# Uncovering hidden Fermi surface instabilities through visualizing unconventional quasi-particle interference in CeTe$_3$

B. R. M. Smith[1*], Y. Fujisawa[1*], P. Wu[1,2], T. Nakamura[1], N. Tomoda[1], S. Kuniyoshi[1,3], D. Ueta[1,4], R. Kobayashi[3], R. Okuma[1,5], K. Arai[5], K. Kuroda[5,6,7], C-H. Hsu[1,8], G. Chang[8], C-Y. Huang[9], H. Lin[9], Z. Y. Wang[2], T. Kondo[5], Y. Okada[1]

* equal contribution

[1]Quantum Materials Science Unit, Okinawa Institute of Science and Technology (OIST), Okinawa 904-0495, Japan.
[2] Department of Physics and Chinese Academy of Sciences Key Laboratory of Strongly-coupled Quantum Matter Physics, University of Science and Technology of China, Hefei, Anhui 230026, China
[3]Faculty of Science, University of the Ryukyus, Nishihara, Okinawa 903-0213, Japan
[4] Institute of Materials Structure Science, High Energy Accelerator Research Organization, Tsukuba, Ibaraki 305-0801, Japan
[5]Institute for Solid State Physics (ISSP), The University of Tokyo, Kashiwa, Chiba 277-8581, Japan
[6]Graduate School of Advanced Science and Engineering, Hiroshima University, Higashi-Hiroshima, Hiroshima 739-8526, Japan
[7]International Institute for Sustainability with Knotted Chiral Meta Matter (WPI-SCKM$^2$), Hiroshima University, Higashi-Hiroshima, Hiroshima 739-8526, Japan
[8]Division of Physics and Applied Physics, School of Physical and Mathematical Sciences, Nanyang Technological University, 637371 Singapore
[9]Institute of Physics, Academia Sinica, Taipei, Taiwan

## Abstract

The charge density wave (CDW) state is a widespread phenomenon in low-dimensional metals/semimetals. The spectral weight of the associated folded bands (shadow bands) can be an intriguing trigger leading to additional Fermi surface instability and unexplored phase transitions. The rare earth tri-telluride CeTe$_3$ exhibits a single CDW stabilized below ~400 K and antiferromagnetism below ~3 K. The distinct periodicities between the Te-square net, the CeTe block layer, and the CDW give rise to rich shadow band formations. In this work, we reveal the predominant scattering between the original and shadow bands at 4 K, with the scattering within the original bands being relatively suppressed at Fermi energy. This unconventional quasi-particle scattering collectively underscores the vital role of the shadow bands' spectral weight and the hidden matrix element effect, which are crucial for controlling electronic properties in this system. Furthermore, our finding points to the existence of rich and unexplored Fermi surface instabilities, which potentially play a role in controlling the nature of long-range antiferromagnetism at lower temperatures in the presence of finite charge-spin interaction.

**Introduction**

Fermi surface (FS) instability represents a fundamental and critical concept in condensed matter physics, giving rise to a diverse range of emergent quantum phases. In scenarios where a phase transition occurs accompanied by FS nesting, the emergence of band folding in k-space is naturally expected. This fundamental picture finds particular relevance in charge density wave (CDW) states commonly observed in low-dimensional metals and semimetals [1]. In many cases of the electronic phase diagram for low-dimensional quantum materials, CDW states exist adjacent to other distinct phases of matter. These include exotic magnetic density wave states, nematic states, excitonic insulators, and unconventional superconductivity [2,3,4,5,6,7]. An essential question is the role of the so-called shadow bands, which originate from band folding with the CDW propagation vector $q_{CDW}$. In principle, the shadow bands can play a role in introducing additional FS instability with a periodicity of $\boldsymbol{q}_{nest}$ [**Fig. 1(a)**]. A powerful approach to investigating this phenomenon is imaging the quasiparticle interference (QPI) using a spectroscopic scanning tunneling microscope (STM). This approach can visualize the degree of the scattering process between the original and the shadow bands, allowing us to find $\boldsymbol{q}_{nest}$ while simultaneously imaging CDW formation in real space.

An excellent materials system to investigate CDW phenomena is the rare earth tri-telluride ($R$Te$_3$) family [**Fig. 1(b, c)**]. The electronic states near the Fermi energy ($E_F$) are dominated by $5p_x$ and $5p_z$ orbitals on the Te-square net [8]. A robust one-dimensional CDW phase is known to exist in the lighter $R$ members, with their characteristic energy scale comparable to room temperature [9,10]. The $R$Te magnetic block layer hosts $4f$-derived long-range antiferromagnetism at low temperatures [11,12,13,14,15,16]. Superconductivity can also be realized under external pressure [17,18]. As seen from the top view [**Fig. 1(c)**], the periodicity for the Te-square net differs from that defined by the crystallographic unit cell [see green square in **Fig. 1(c)**]. Consequently, folding of the original $5p_x$ and $5p_z$ bands due to the crystallographic unit cell creates shadow bands. We will refer to these bands as $q_{cry}$-shadow bands hereafter. These $q_{cry}$-shadow bands are additional to those arising from the CDW shadow bands (referred to as $q_{CDW}$-shadow bands hereafter). In this letter, we shall refer to both the $q_{cry}$-shadow bands and the $q_{CDW}$-shadow bands together in the term "shadow bands" to host possible triggers for exotic FS instability. Hence, a deeper understanding of the nature and interactions of the multiple shadow bands is a crucial foundation for revealing the mechanisms underlying various emergent electronic and magnetic phases in $R$Te$_3$.

It has been shown that the inherently low-dimensional and exfoliable nature of the $R$Te$_3$ makes it suitable for single particle spectroscopy such as angle-resolved photoemission spectroscopy (ARPES) and STM

[19,20,21,22,23,24,25,26,27,28,29]. Although this system presents an excellent platform for investigating the role of shadow bands through QPI imaging, such studies are conspicuously absent from the literature. This paper addresses this gap using $CeTe_3$ as a model system. We demonstrate the observation of unconventional QPI signals involving the significant contribution of shadow bands' state, while the scattering between the original bands is relatively suppressed. This observation suggests the presence of intriguing and hidden Fermiology in this system.

## Results and Discussions

Cleavage along the van der Waals gap reliably exposes the Te square net observed in our STM topography. A typical example of a topographic image for a relatively wide area (120×120 nm) is shown in **Fig. 1(d)**, with a magnified atomic resolution image inset (see also [30] for other images). The 2D fast Fourier transformation (FFT) of **Fig. 1(d)** is shown in **Fig. 1(e)**. We observe the Bragg peaks from Te atoms in the Te square net (red square) and those from the crystallographic unit cell periodicity labelled as $q_c$ and $q_a$ (black circles). We can determine the nearest-neighbor Te bond length on the square net from these Bragg points as 3.1 Å. This value is consistent with bulk XRD characterization [31]. The peaks from the unidirectional incommensurate CDW are also visible along the $q_c$ direction (blue circle). The primary CDW wavevector can be estimated as $q_{CDW}=5/7 q_c$, which is evident as a charge modulation in real space (inset of **Fig. 1(d)**). This value agrees with literature reports [24,25,26]. As is also discussed in the literature, since the secondary CDW propagation vector $q'_{CDW}=q_c-q_{CDW}$ is allowed, various reflections and combinations of $q_{CDW}$ and $q'_{CDW}$ are observed in the FFT image (see also **Fig 1(f)** for the line profile along the black arrow in **Fig. 1(e)**). The spatially averaged $dI/dV$ spectrum and its derivative are shown in **Fig. 1(g)**, with apparent anomalies near the $E_F$ (see two dashed vertical lines). The large energy scale (~ 550 meV) can be assigned as the CDW gap consistent with the previous spectroscopic studies in $CeTe_3$ [25,26,32].

We present our observation of the QPI patterns near $E_F$ in the range ±30 meV. The $dI/dV$ maps at -30, 0, and +30 meV are shown in **Fig. 2(a,b,c)**, taken in the same area as in **Fig. 1(d)**. The wave-like conductance corrugations are observed in the real space images. FFT images of the conductance maps are shown in **Fig. 2(d,e,f)** to extract the characteristic wave vectors. The streak-like pattern is displayed clearly for all energies [**Fig. 2(d,e,f)**]. Here, FFT images around $q$ = 0 are selectively displayed, as the other signals can also be assigned by the representative signals around $q$ = 0 (see [30] for the full assignment of the QPI channels). For clarity, the three relevant streak patterns are overlaid with colored lines labelled as $q_{s1}$, $q_{s2}$, and $q_{s3}$ [**Fig. 2(d)**]. To display the energy dependence of these streaks, the energy dispersion linecuts taken along the black arrow in **Fig. 2(e)** are shown in **Fig. 2(g)**. The dispersion lines for $q_{s2}$ and $q_{s3}$ were plotted by peak fitting on the $q$-dependent intensity profile. While the slope is steep, we confirm they are dispersive; $q_{s2}$ has electron-like dispersion, and $q_{s3}$ has hole-like dispersion. On the other hand, because of the low intensity of the $q_{s1}$ channel compared to background noise impeding a reliable fit, the line for $q_{s1}$ is a simple guide to the eye based on the color contrast in **Fig. 2(g)**. As demonstrated later, the scattering channels for $q_{s1}$, $q_{s2}$, and $q_{s3}$ are interpreted collectively as QPI origin [33,34,35,36].

To simulate the QPI patterns, the relevant bands are considered by solving a 12x12 Hamiltonian matrix (see [30] for details) [19,20,23]. We begin with a tight-binding description of the original $5p_x$ and $5p_z$ bands [**Fig. 3(a)**]. Based on these two parent bands, we then model the $q_{cry}$-shadow bands [**Fig. 3(b)**] and the $q_{CDW}$-shadow bands [**Fig. 3(c)**]. $q_{cry}$ can be readily determined from the crystal symmetry [**Fig. 1(c)**], and $q_{CDW}$ is determined experimentally [**Fig. 1(e)**]. While the bands shown in **Fig. 3(a-c)** are non-interacting, we implement finite off-diagonal terms in the Hamiltonian matrix to get interacting FS shown in **Fig. 3(d)**. **Figure 3(e)** demonstrates the agreement between our model and SX-ARPES data. Note that the main ARPES intensity is derived from the original band [black line in **Fig. 3(d,e)**], and the intensity from the shadow bands is relatively low. This is due to the smaller spectral weight of the shadow bands, because of the smaller coupling between the electrons and the superperiodicities [19, 23]. Despite this lower intensity, the QPI imaging demonstrates the surprisingly significant role of quasiparticles on the shadow bands.

The QPI pattern is simulated using the joint density of states (JDOS) approximation [35,36]. We reasonably focus on the pattern at $E_F$ due to the steep slope of the band dispersion, leading to an absence of dramatic energy evolution of the QPI patterns in the range studied. To deepen understanding of the $k$-space origin of our QPI signal, using the FS shown in **Fig. 3(d,e)**, three cases of the JDOS are simulated (see [30] for matrix representation of the considered scattering channels for each case). The first is the JDOS purely from the original two bands [**Fig. 4(a)**], the second is the JDOS of the original bands and the $q_{cry}$-shadow bands [**Fig. 4(b)**], and the third is the JDOS of the original bands and the $q_{CDW}$-shadow bands [**Fig. 4(c)**]. In our simulations, we consider the scatterings only between the bands of the same orbital character (i.e., between $p_x$ bands and between $p_z$ bands) and disregard the scatterings between the bands of different orbital characters (i.e., between $p_x$ bands and $p_z$ bands) because they contribute to a $q$-independent background in the JDOS as shown in [30].

In the JDOS simulation based on the original two bands [**Fig. 4(a)**], while $q_{s1}$ is reproduced in $q$-space, the experimentally observed $q_{s2}$ and $q_{s3}$ are not fully reproduced [see **Fig. 2(d)**]. Since this simulation does not exclude any possible scattering through the matrix element effect, considering the scattering involving the shadow bands is required. Indeed, after considering the $q_{cry}$ (and $q_{CDW}$) -shadow bands, the $q_{s2}$ ($q_{s3}$) channels are reproduced [see **Fig. 4(b,c)**]. The three scattering channels ($q_{s1}$, $q_{s2}$, and $q_{s3}$), which should connect two distinct Fermi surface portions, are mapped in k-space to present the momentum origin of scattering channels [see the red arrows in **Fig. 4(d), (e), and (f)**]. As in **Fig. 4(g)**, the consistency of this assignment, beyond $E_F$, is further checked by comparing the energy dispersion of $q_{s2}$ and $q_{s3}$ with band dispersion along the black arrow in **Fig. 4(d)**. The inner Fermi surface in which $q_{s2}$ occurs is electron-like, whereas the inter-pocket Fermi surface portion in which $q_{s3}$ occurs has hole-like dispersion. These dispersion characters are consistent with the experiments [**Fig. 2(g)**].

To explain experimental $q_{s1}$, $q_{s2}$, and $q_{s3}$ collectively, it is naturally motivated to compare the experimental QPI pattern with the JDOS simulation, which integrates all three scattering cases [**Fig. 4(a)-(c)**]. As shown in **Fig. 4(h)**, the comparison represents a good agreement. The most crucial point is that the experimental QPI signals originating from the scatterings within the original bands ($q_{s1}$) have a lower intensity than those between the original and the shadow bands ($q_{s2}$ and $q_{s3}$). While further theoretical and experimental investigations are necessary to establish the physics behind this firmly, we propose the existence of a hidden matrix element and potential unconventional electronic texture on the Fermi surface (see [30] for the detailed discussion). Furthermore, the observed QPI signals point to the dominant role of the spectral weight of the shadow bands determining near $E_F$ electronic nature in the system. This also allows us to consider a rich landscape of competing Fermi surface nesting $q_{nest}$ at the zone boundary under the existence of the robust CDW with $q_{CDW}$ [see **Fig. 4(d-f)**].

It is particularly intriguing to discuss how the multiplicity of the Fermi surface instability represented by $q_{nest}$ affects the lower temperature magnetic phases in $R$Te$_3$ [37,38,39,40]. Because the intimate interplay between the conduction electrons and localized 4$f$ spin moment is proposed as in recent studies [13,14,15], it is possible that the emergence of magnetic phases gives rise to the reduction of the total electron energy by the formation of the new charge orders with $q_{nest}$. Within the Kondo lattice model, the magnetic ordering vectors ($q_m$) can be described either by $q_m = q_{nest}$ or $q_m = q_{nest}/2$, depending on whether there is a finite net moment or not [41,42,43]. Since the nearly commensurate magnetic vectors are experimentally observed in the other $R$Te$_3$ systems [13,14,15], the existence of commensurate $q_{nest}$ is naturally expected. Therefore, to focus on scattering channels with relatively enhanced charge susceptibility, we mainly shed light on the commensurate $q_{nest}$ in the QPI signal at the $E_F$ [see **Fig. 4(i)**] [44]. As the red circle indicates, one potential ordering vector $q_{nest}$ is (0,1/3). This wave vector is well known as the wave vector of the secondary CDW in heavier $R$Te$_3$. While the secondary CDW is absent in CeTe$_3$ above 4 K, it has been shown to emerge under non-equilibrium conditions, suggesting its inherent instability in CeTe$_3$ [9]. Therefore, a single-$q$ magnetic structure, such as $q_m$=(0,1/3) and (0,1/6) may emerge associated with this instability at lower temperatures in CeTe$_3$. Another possible $q_{nest}$ is the equivalent (1/5,1/5) and (1/5,-1/5) shown as blue circles in **Fig. 4(i)**. These $q_{nest}$ vectors are unique because they collectively satisfy commensurability, four-fold symmetry, and significant intensity in the QPI. Therefore, the magnetic structure with $q_m$=(1/5,±1/5) and (1/10, ±1/10) is an interesting candidate. For the future, searching potential topological non-trivial phases can be an interesting subject, as has been explored in different skyrmion materials [42,43]. Note that the Fermi surface instability led by $q_{CDW}$ cannot solely cause these magnetic transitions, considering the high CDW transition temperature (~400 K) and the low magnetic transition temperatures (<3 K). We believe that the interplay between charge, phonons [45], and 4$f$ spins is important for the emergent magnetic phases, in addition to the Fermi surface instability revealed in this study [13,14,15].

In summary, we report the QPI visualization of the prototypical CDW material CeTe$_3$ for the first time. We found a surprising enhancement of the QPI signals arising from scattering between the original bands and the shadow bands, with the scattering within the original bands being relatively suppressed. This indicates the existence of a hidden electronic texture and associated matrix element on the Fermi surface in CeTe$_3$. Furthermore, the coupling between the electrons and 4$f$ spins enables the emergence of enriched Fermi surface instability, which potentially plays a vital role in leading to intriguing unexplored long-range antiferromagnetism below 3 K. Because QPI is the only way to resolve electronic deformation under a magnetic field with high energy and real space resolution, our first report of QPI will be useful to understand the intimate connection between Fermi surface instability and magnetic phases in $R$Te$_3$.

## Acknowledgment

We thank Y. Kohsaka, T. Machida, and C.Y. Huang for fruitful discussions. Crystal structures were visualized using Vesta [46].

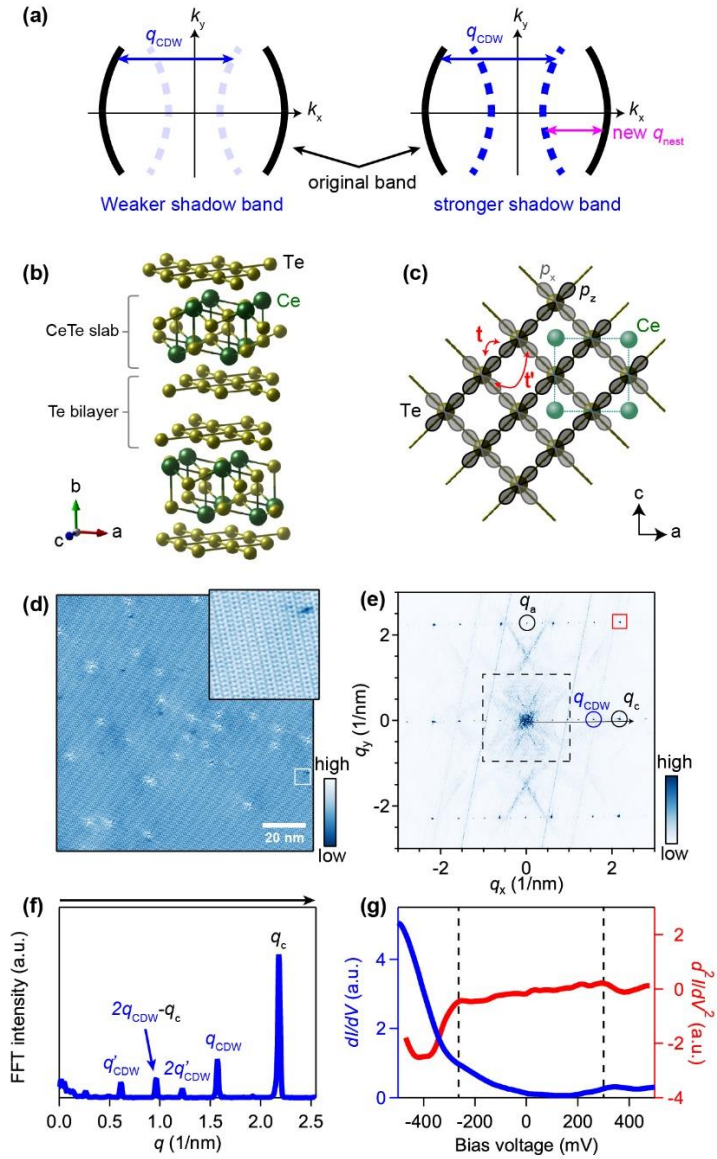

**Fig. 1.**

(a) Schematic representation of the Fermi surface with CDW shadow bands with lower intensity (left) and higher intensity (right). (b) The crystal structure of $CeTe_3$. (c) Top view of the top-most Te square net, with the $p_x$ and $p_z$ orbital chains used for the tight binding calculation. The square lattice of Ce atoms is also indicated with green atoms. (d) STM topography of the $CeTe_3$ surface with a feedback condition of 30 mV/500 pA. The inset is a close-up of the white box. (e) An FFT image of d after rotation for clarity. See the main text for each symbol in the image. (f) The line cut along the black arrow is in e. (g) A typical averaged $dI/dV$ spectrum (left axis) and its derivative (right axis). The black dashed lines indicate the CDW gap. See [30] for the detailed experimental condition.

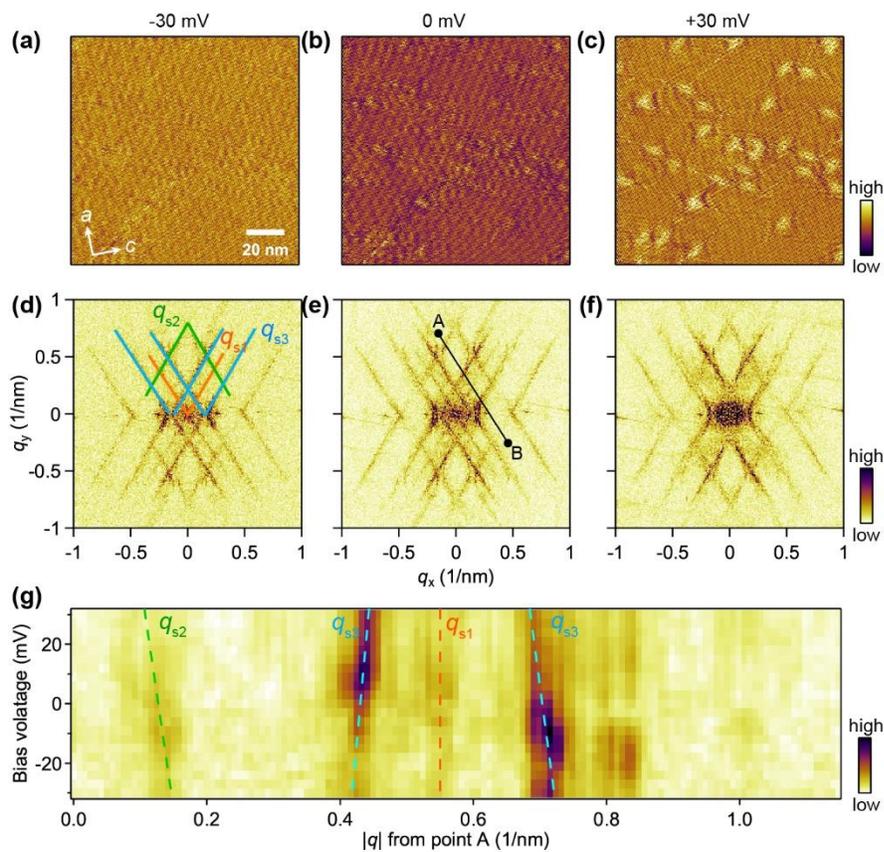

**Fig. 2.**

(a-c) *dI/dV* maps taken on the same surface are shown in **Fig. 1d** at -30 meV, 0 meV, and +30 meV, respectively. See [30] for the detailed experimental condition.

(d-f) corresponding FFT images of the *dI/dV* maps. In (d), the colored lines are guides to the eye, indicating the main scattering channels.

(g) Energy dispersions of the observed QPI pattern along the black arrow in (e).

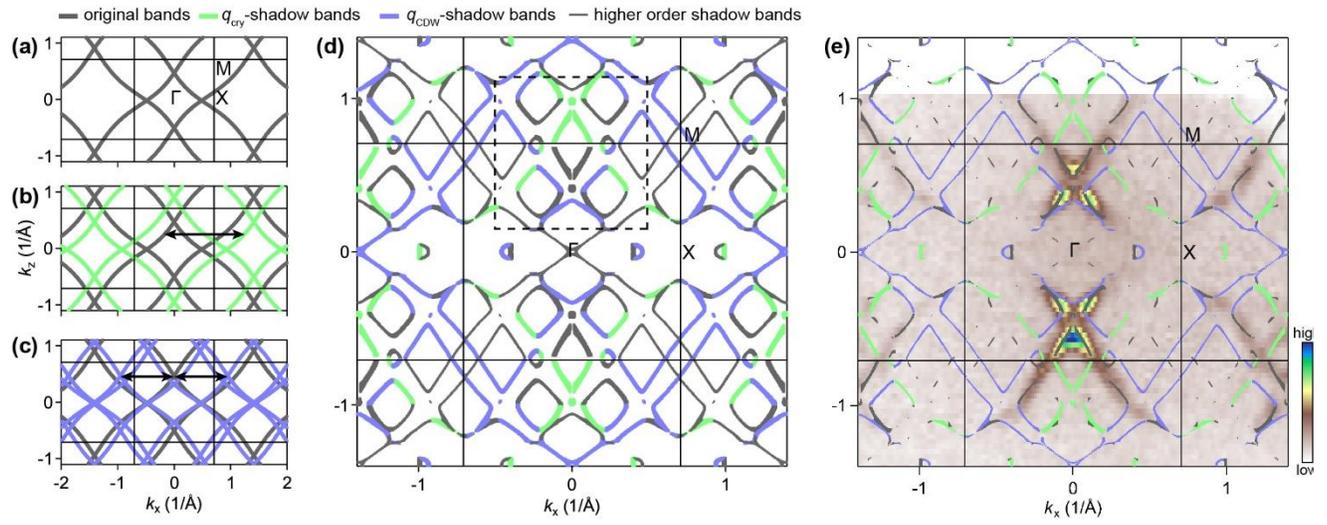

**Fig. 3.**

(a-c) Non-interacting Fermi surface (FS) considering the original $p_x$ and $p_z$ (a), the original bands and the $q_{cry}$-shadow bands (b), and the original bands and the $q_{CDW}$-shadow bands (c). The black arrows in b and c represent $q_{cry}$ and $q_{CDW}$, respectively. The black squares indicate the BZ for the crystallographic unit cell.

(d) Interacting FS considering the original, the $q_{cry}$-shadow, the $q_{CDW}$-shadow, and the higher order shadow bands that are $q_{CDW}$-shadow bands folded by $q_{cry}$.

(e) SX-ARPES derived FS showing agreement with the simulated Femi surface contributing from the original bands. Energy integration is 80 meV below the $E_F$.

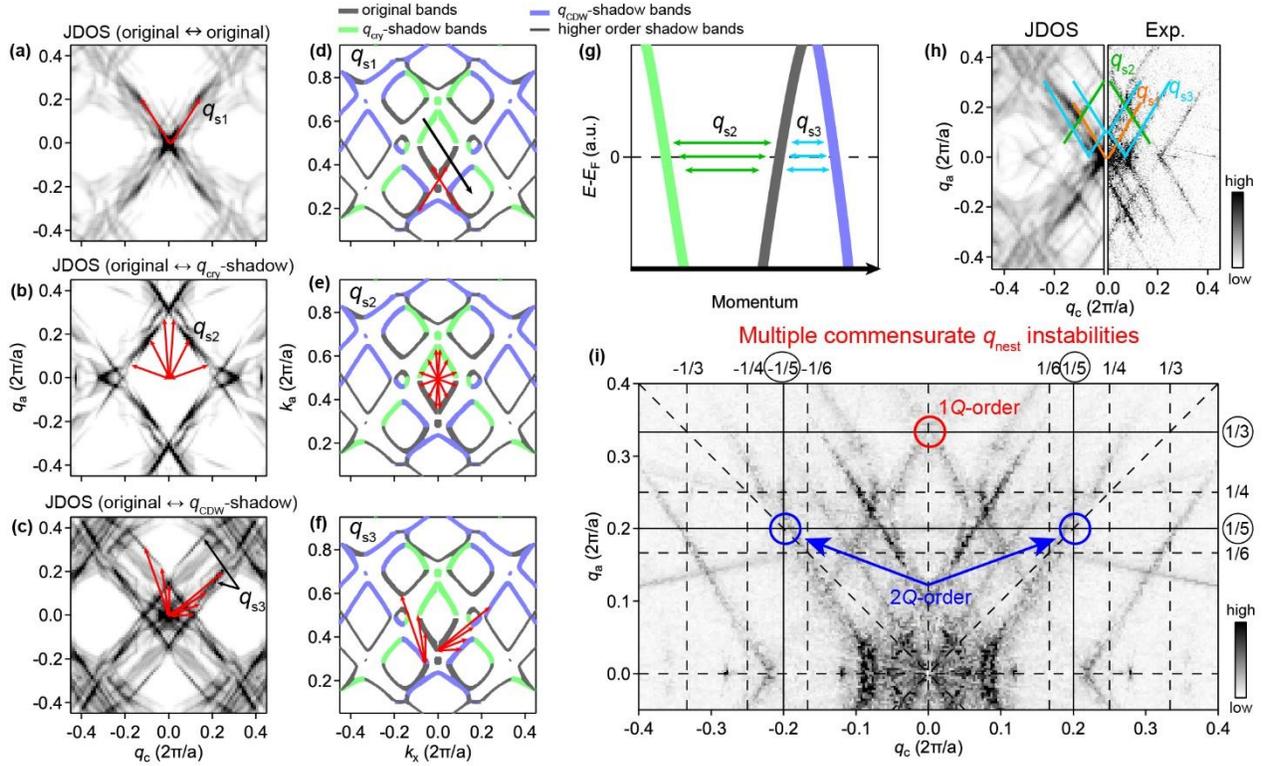

**Fig 4.**

(a-c) JDOS simulation considering only the original bands (a), the original bands and the $q_{cry}$-shadow bands (b), and the original and the $q_{CDW}$-shadow bands (c). The red arrows indicate the representative scattering vectors.

(d-f) Tight-binding-derived Fermi surface. The red arrows represent the possible scattering channels for $q_{s1}$ (d), $q_{s2}$ (e), and $q_{s3}$ (f), respectively.

(g) Tight-binding-derived energy dispersion along the black arrow in (d).

(h) Comparison of the QPI pattern obtained from JDOS considering the original, the $q_{cry}$-shadow, and the $q_{CDW}$-shadow bands (left) and experiments (right).

(i) Experimentally obtained QPI data at the Fermi level, highlighting the existence of multiple Fermi surface instabilities generated by the CDW with $q_{CDW}$.

Therefore, we safely ignore the energy-dependent $q_{nest}$ due to the steep dispersion.